\documentclass[twocolumn]{aastex61}

\usepackage{xspace}
\usepackage{upgreek}
\usepackage{color}

\newcommand{\frb}{FRB~121102\xspace}

\newcommand{\msun}{\mathrm{M_{\odot}}\xspace}
\newcommand{\degree}{^{\circ}\xspace}

\received{2016 December 10}
\revised{2016 December 21}
\accepted{2016 December 22}
\submitjournal{ApJ}

\shorttitle{FRB~121102 as Seen on Milliarcsecond Angular Scales}
\shortauthors{Marcote et al.}

\begin{document}

\title{The Repeating Fast Radio Burst FRB~121102 as Seen on Milliarcsecond Angular Scales}

\correspondingauthor{J.~W.~T.~Hessels}
\email{J.W.T.Hessels@uva.nl, marcote@jive.eu}

\author[0000-0001-9814-2354]{B.~Marcote}
\affiliation{Joint Institute for VLBI ERIC, Postbus 2, 7990 AA Dwingeloo, The Netherlands}

\author{Z.~Paragi}
\affiliation{Joint Institute for VLBI ERIC, Postbus 2, 7990 AA Dwingeloo, The Netherlands}

\author[0000-0003-2317-1446]{J.~W.~T.~Hessels}
\affiliation{ASTRON, Netherlands Institute for Radio Astronomy, Postbus 2, 7990 AA Dwingeloo, The Netherlands}
\affiliation{Anton Pannekoek Institute for Astronomy, University of Amsterdam, Science Park 904, 1098 XH Amsterdam, The Netherlands}

\author{A.~Keimpema}
\affiliation{Joint Institute for VLBI ERIC, Postbus 2, 7990 AA Dwingeloo, The Netherlands}

\author{H.~J.~van~Langevelde}
\affiliation{Joint Institute for VLBI ERIC, Postbus 2, 7990 AA Dwingeloo, The Netherlands}
\affiliation{Sterrewacht Leiden, Leiden University, Postbus 9513, 2300 RA Leiden, The Netherlands}

\author{Y.~Huang}
\affiliation{Department of Physics and Astronomy, Carleton College, Northfield, MN 55057, USA}
\affiliation{Joint Institute for VLBI ERIC, Postbus 2, 7990 AA Dwingeloo, The Netherlands}

\author{C.~G.~Bassa}
\affiliation{ASTRON, Netherlands Institute for Radio Astronomy, Postbus 2, 7990 AA Dwingeloo, The Netherlands}

\author{S.~Bogdanov}
\affiliation{Columbia Astrophysics Laboratory, Columbia University,  New York, NY 10027, USA}

\author{G.~C.~Bower}
\affiliation{Academia Sinica Institute of Astronomy and Astrophysics, 645 N. A'ohoku Place, Hilo, HI 96720, USA}

\author{S.~Burke-Spolaor}
\affiliation{National Radio Astronomy Observatory, Socorro, NM 87801, USA}
\affiliation{Department of Physics and Astronomy, West Virginia University, Morgantown, WV 26506, USA}
\affiliation{Center for Gravitational Waves and Cosmology, West Virginia University, Chestnut Ridge Research Building, Morgantown, WV 26505, USA}

\author{B.~J.~Butler}
\affiliation{National Radio Astronomy Observatory, Socorro, NM 87801, USA}

\author{R.~M.~Campbell}
\affiliation{Joint Institute for VLBI ERIC, Postbus 2, 7990 AA Dwingeloo, The Netherlands}

\author{S.~Chatterjee}
\affiliation{Cornell Center for Astrophysics and Planetary Science and Department of Astronomy, Cornell University, Ithaca, NY 14853, USA}

\author{J.~M.~Cordes}
\affiliation{Cornell Center for Astrophysics and Planetary Science and Department of Astronomy, Cornell University, Ithaca, NY 14853, USA}

\author{P.~Demorest}
\affiliation{National Radio Astronomy Observatory, Socorro, NM 87801, USA}

\author{M.~A.~Garrett}
\affiliation{Jodrell Bank Centre for Astrophysics, University of Manchester, M13 9PL, UK}
\affiliation{Sterrewacht Leiden, Leiden University, Postbus 9513, 2300 RA Leiden, The Netherlands}
\affiliation{ASTRON, Netherlands Institute for Radio Astronomy, Postbus 2, 7990 AA Dwingeloo, The Netherlands}

\author{T.~Ghosh}
\affiliation{Arecibo Observatory, HC3 Box 53995, Arecibo, PR 00612, USA}

\author{V.~M.~Kaspi}
\affiliation{Department of Physics and McGill Space Institute, McGill University, 3600 University St., Montreal, QC H3A 2T8, Canada}

\author{C.~J.~Law}
\affiliation{Department of Astronomy and Radio Astronomy Lab, University of California, Berkeley, CA 94720, USA}

\author{T.~J.~W.~Lazio}
\affiliation{Jet Propulsion Laboratory, California Institute of Technology, Pasadena, CA 91109, USA}

 \author{M.~A.~McLaughlin}
 \affiliation{Department of Physics and Astronomy, West Virginia University, Morgantown, WV 26506, USA}
 \affiliation{Center for Gravitational Waves and Cosmology, West Virginia University, Chestnut Ridge Research Building, Morgantown, WV 26505, USA}

\author{S.~M.~Ransom}
\affiliation{National Radio Astronomy Observatory, Charlottesville, VA 22903, USA}

\author{C.~J.~Salter}
\affiliation{Arecibo Observatory, HC3 Box 53995, Arecibo, PR 00612, USA}

\author{P.~Scholz}
\affiliation{National Research Council of Canada, Herzberg Astronomy and Astrophysics, Dominion Radio Astrophysical Observatory, P.O. Box 248, Penticton, BC V2A 6J9, Canada}

\author{A.~Seymour}
\affiliation{Arecibo Observatory, HC3 Box 53995, Arecibo, PR 00612, USA}

\author{A.~Siemion}
\affiliation{Department of Astronomy and Radio Astronomy Lab, University of California, Berkeley, CA 94720, USA}
\affiliation{ASTRON, Netherlands Institute for Radio Astronomy, Postbus 2, 7990 AA Dwingeloo, The Netherlands}
\affiliation{Radboud University, 6525 HP Nijmegen, The Netherlands}

\author{L.~G.~Spitler}
\affiliation{Max-Planck-Institut f\"ur Radioastronomie, Auf dem H\"ugel 69, Bonn, D-53121, Germany}

\author{S.~P.~Tendulkar}
\affiliation{Department of Physics and McGill Space Institute, McGill University, 3600 University St., Montreal, QC H3A 2T8, Canada}

\author{R.~S.~Wharton}
\affiliation{Cornell Center for Astrophysics and Planetary Science and Department of Astronomy, Cornell University, Ithaca, NY 14853, USA}

\begin{abstract}

The millisecond-duration radio flashes known as Fast Radio Bursts (FRBs) represent an enigmatic astrophysical phenomenon.  Recently, the sub-arcsecond localization ($\sim 100$~mas precision) of \frb using the VLA has led to its unambiguous association with persistent radio and optical counterparts, and to the identification of its host galaxy. However, an even more precise localization is needed in order to probe the direct physical relationship between the millisecond bursts themselves and the associated persistent emission.  Here we report very-long-baseline radio interferometric observations using the European VLBI Network and the 305-m Arecibo telescope, which simultaneously detect both the bursts and the persistent radio emission at milliarcsecond angular scales and show that they are co-located to within a projected linear separation of $\lesssim 40$~pc ($\lesssim 12$~mas angular separation, at 95\% confidence).  We detect consistent angular broadening of the bursts and persistent radio source ($\sim 2$--$4~\mathrm{mas}$ at 1.7~GHz), which are both similar to the expected Milky Way scattering contribution. The persistent radio source has a projected size constrained to be $\lesssim 0.7~\mathrm{pc}$ ($\lesssim 0.2$~mas angular extent at 5.0~GHz) and a lower limit for the brightness temperature of $T_b \gtrsim 5 \times 10^7\ \mathrm{K}$.  Together, these observations provide strong evidence for a direct physical link between \frb and the compact persistent radio source. We argue that a burst source associated with a low-luminosity active galactic nucleus or a young neutron star energizing a supernova remnant are the two scenarios for \frb that best match the observed data.

\end{abstract}


\section{Introduction} \label{sec:intro}

Fast Radio Bursts (FRBs) are transient sources of unknown physical origin, which are characterized by their short ($\sim$~ms), highly dispersed, and bright ($S_{\rm peak} \sim 0.1$--$10~\mathrm{Jy}$) pulses. Thus far, 18 FRBs have been discovered using single dish observations \citep[e.g.,][]{lbm+07,tsb+13,pbj+16}. Unambiguous associations with multiwavelength counterparts have been extremely limited by the poor localization that such telescopes provide (uncertainty regions of at least several square arcminutes). \citet{keane2016} reported the first apparent localization of an FRB by associating FRB~150418 with a pseudo-contemporaneous transient radio source in a galaxy at $z \sim 0.5$. However, further studies have shown that the transient source continues to vary in brightness well after the initial FRB~150418 burst detection, and can be explained by a scintillating, low-luminosity active galactic nucleus (AGN; e.g. \citealt{wb16,gmg+16,bassa2016,johnston2016}), which leaves limited grounds to claim an unambiguous association with FRB~150418.

Thus far, \frb is the only known FRB to have shown repeated bursts with consistent dispersion measure (DM) and sky localization \citep{sch+14,ssh+16a,ssh+16b}. Recently, using fast-dump interferometric imaging with the Karl G. Jansky Very Large Array (VLA), \frb has been localized to $\sim 100$~milliarcsecond (mas) precision \citep{chatterjee2017}. The precise localization of these bursts has led to associations with both persistent radio and optical sources, and the identification of \frb's host galaxy \citep{chatterjee2017,tendulkar2017}.  European VLBI Network (EVN) observations, confirmed by the Very Long Baseline Array (VLBA), have shown that the persistent source is compact on milliarcsecond scales \citep{chatterjee2017}.  Optical observations have identified a faint ($m_{r^{\prime}} = 25.1 \pm 0.1$~AB~mag) and extended ($0.6$--$0.8~\mathrm{arcsec}$) counterpart in Keck and Gemini data, located at a redshift $z = 0.19273 \pm 0.00008$ -- i.e. at a luminosity distance of $D_{\rm L} \approx 972\ \mathrm{Mpc}$, and implying an angular diameter distance of $D_{\rm A} \approx 683\ \mathrm{Mpc}$ \citep{tendulkar2017}. The centroids of the persistent optical and radio emission are offset from each other by $\sim 0.2~\mathrm{arcsec}$, and the observed optical emission lines are dominated by star formation, with an estimated star formation rate of $\sim 0.4\ \msun\ \mathrm{yr}^{-1}$ \citep{tendulkar2017}.  In X-rays, {\em XMM-Newton} and {\em Chandra} observations provide a 5-$\sigma$ upper limit in the $0.5$--$10$~keV band of $L_X < 5 \times 10^{41}\ \mathrm{erg\ s^{-1}}$ \citep{chatterjee2017}.

In the past few years, significant efforts have been made to detect and localize millisecond transient signals using the EVN \citep{paragi2017}. This was made possible by the recently commissioned EVN Software Correlator (SFXC; \citealt{kkp+15}) at the Joint Institute for VLBI ERIC (JIVE; Dwingeloo, the Netherlands).  Here we present joint Arecibo and EVN observations of \frb which simultaneously detect both the persistent radio source as well as four bursts from \frb, localizing both to milliarcsecond precision. In \S\ref{sec:obs} we present the observations and data analysis. In \S\ref{sec:results} we describe the results and in \S\ref{sec:discussion} we discuss the properties of the persistent source and its co-localization with the source of the bursts.  A discussion of the constraints that these data place on the physical scenarios is also provided. Finally, we present our conclusions in \S\ref{sec:conclusion}.

\section{Observations and Data Analysis} \label{sec:obs}

We have observed \frb using the EVN at 1.7~GHz and 5~GHz central frequencies (with a maximum bandwidth of 128~MHz in both cases) in 8 observing sessions that span 2016 Feb 1 to Sep 21 (Table~\ref{tab:data}). These observations included the 305-m William E. Gordon Telescope at the Arecibo Observatory (which provides raw sensitivity for high signal-to-noise burst detection) and the following regular EVN stations: Effelsberg, Hartebeesthoek, Lovell Telescope or Mk2 in Jodrell Bank, Medicina, Noto, Onsala, Tianma, Toru\'{n}, Westerbork (single dish), and Yebes.  Of these antennas, Hartebeesthoek, Noto, Tianma, and Yebes only participated in the single 5-GHz session.

\begin{deluxetable*}{cccr@{ }c@{ }lr@{ }c@{ }lr@{ }c@{ }lc}
	\tablecaption{Properties of the persistent radio source and detected \frb bursts from the Arecibo+EVN observations.  All positions are referred to the 5-GHz detection of the persistent source (RP026C epoch): $\alpha_{\rm J2000} = 5^{\rm h}31^{\rm m}58.70159^{\rm s}$, $\delta_{\rm J2000} = 33^{\circ}8^{\prime}52.5501^{\prime\prime}$.  The observations conducted on 2016 Feb 1 (RP024A) and 2016 Sep 19 (RP026A) did not produce useful data, and are not included here (see main text). The arrival times of the bursts are UTC topocentric at Arecibo at the top of the observing band (1690.49~MHz).  All these bursts had gate widths of $2$--$3$~ms, and the quoted flux densities are averages over these time windows.  We note that the larger error on the flux density of Burst~\#2 is due to the fact that the image is dynamic-range limited because of the burst's brightness.  The last row shows the average position obtained from the four bursts weighted by the detection statistic $\xi = F/\sqrt{w}$ (fluence divided by the square-root of the burst width).
	\label{tab:data}}
	\tablehead{
		\colhead{Session}&\colhead{Epoch}&\colhead{$\nu$}&\multicolumn{3}{c}{$\Delta\alpha$}& \multicolumn{3}{c}{$\Delta\delta$}&\multicolumn{3}{c}{$S_{\nu}$}&\colhead{$\xi$} \\
		&\colhead{(YYYY-MM-DD)}&\colhead{(GHz)}& \multicolumn{3}{c}{(mas)}&\multicolumn{3}{c}{(mas)}&\multicolumn{3}{c}{$\mathrm{(\upmu Jy)}$}&\colhead{(Jy ms$^{1/2}$)}
	}
	\startdata
		RP024B & 2016-02-10 & $1.7$ & $1.5 $&$\pm$&$ 2$ & $-2 $&$\pm$&$ 3$ & $200 $&$\pm$&$ 20$&--- \\
		RP024C & 2016-02-11 & $1.7$ & $-4 $&$\pm$&$ 2$ & $-5 $&$\pm$&$ 3$ & $175 $&$\pm$&$ 14$&--- \\
		RP024D & 2016-05-24 & $1.7$ & $1 $&$\pm$&$ 3$ & $-5 $&$\pm$&$ 4$ & $220 $&$\pm$&$ 40$&--- \\
		RP024E & 2016-05-25 & $1.7$ & $1 $&$\pm$&$ 3$ & $2 $&$\pm$&$ 4$ & $180 $&$\pm$&$ 40$&--- \\
		RP026B & 2016-09-20 & $1.7$ & $1.9 $&$\pm$&$ 1.8$ & $-0.4 $&$\pm$&$ 2.3$ & $168 $&$\pm$&$ 11$&--- \\
		RP026C & 2016-09-21 & $5.0$ & $0.0 $&$\pm$&$ 0.6$ & $0.0 $&$\pm$&$ 0.7$ & $123 $&$\pm$&$ 14$&--- \\
		\hline
				& (YYYY-MM-DD hh:mm:ss.sss) & &&&&& & & \multicolumn{3}{c}{(Jy)} \\
		\hline
		Burst \#1 & 2016-09-20 09:52:31.634 & $1.7$ & $-14 $&$\pm$&$ 3$ & $-1.4 $&$\pm$&$ 1.8$ & $0.46 $&$\pm$&$ 0.02$&$\sim 0.8$ \\
		Burst \#2 & 2016-09-20 10:02:44.716 & $1.7$ & $-3.3 $&$\pm$&$ 2.5$ & $4.3 $&$\pm$&$ 1.6$ & $3.72 $&$\pm$&$ 0.12$&$\sim 5$ \\
		Burst \#3 & 2016-09-20 10:03:29.590 & $1.7$ & $-10 $&$\pm$&$ 5$ & $0.8 $&$\pm$&$ 3$ & $0.22 $&$\pm$&$ 0.03$&$\sim 0.4$ \\
		Burst \#4 & 2016-09-20 10:50:57.695 & $1.7$ & $3 $&$\pm$&$ 6$ & $6 $&$\pm$&$ 4$ & $0.17 $&$\pm$&$ 0.03$&$\sim 0.2$ \\
		Avg. burst pos. & 2016-09-20 & $1.7$ & $-5 $&$\pm$&$ 4$ & $3.5 $&$\pm$&$ 2.2$ & \multicolumn{3}{c}{---}&--- \\
	\enddata
\end{deluxetable*}

We simultaneously acquired both EVN VLBI data products (buffered baseband data and real-time correlations) as well as wideband, high-time-resolution data from Arecibo as a stand-alone telescope. The Arecibo single-dish data provide poor angular resolution ($\sim 3~\mathrm{arcmin}$ at 1.7~GHz), but unparalleled sensitivity in order to search for faint millisecond bursts. By first detecting bursts in the Arecibo single-dish data, we could then zoom-in on specific times in the multi-telescope EVN data set where we could perform high-angular-resolution imaging of the bursts themselves.

\subsection{Arecibo Single Dish Data}

For the 1.7-GHz observations, Arecibo single-dish observations used the Puerto-Rican Ultimate Pulsar Processing Instrument (PUPPI) in combination with the L-band Wide receiver, which provided $\sim 600~\mathrm{MHz}$ of usable bandwidth between $1150$--$1730~\mathrm{MHz}$. The PUPPI data were coherently dedispersed to a $\mathrm{DM} = 557\ \mathrm{pc\ cm^{-3}}$, as previously done by \citet{ssh+16b}. Coherent dedispersion removes the dispersive smearing of the burst width within each spectral channel. The time resolution of the data was $10.24\,\mathrm{\upmu s}$, and we recorded full Stokes parameters.  At 5~GHz, the Arecibo single-dish observations were recorded with the Mock Spectrometers in combination with the C-band receiver, which together provided spectral coverage from $4484$--$5554~\mathrm{MHz}$. The Mock data were recorded in 7 partially overlapping subbands of 172~MHz, with 5.376-MHz channels and $65.476~\mathrm{\upmu s}$ time resolution. In addition to the PUPPI and Mock data, the autocorrelations of the Arecibo data from the VLBI recording were also available (these are restricted to only 64~MHz of bandwidth, see below).

The Arecibo single-dish data were analyzed using tools from the {\tt PRESTO}\footnote{Available at https://github.com/scottransom/presto} suite of pulsar software \citep{ran01}, and searched for bursts using standard procedures \citep[e.g.,][]{ssh+16b}.  The data were first subbanded to $8\times$ lower time and frequency resolution and were then dedispersed using {\tt prepsubband} to trial DMs between $487$--$627\ \mathrm{pc\,cm^{-3}}$ in order to search for pulses that peak in signal-to-noise ratio (S/N) at the expected DM of \frb. This is required to separate astrophysical bursts from radio frequency interference (RFI).  For each candidate burst found using {\tt single\_pulse\_search.py} (and grouping common events across DM trials), the astrophysical nature was confirmed by producing a frequency versus time diagram to show that the signal is (relatively) broadband compared to the narrow-band RFI signals that can sometimes masquerade as dispersed pulses.

\subsection{Arecibo+EVN Interferometric Data}

EVN data were acquired in real time using the e-EVN setup, in which the data are transferred to the central processing center at JIVE via high-speed fibre networks and correlated using the SFXC software correlator. The high data rate of VLBI observations requires visibilities to be typically averaged to 2-s intervals during correlation, which is sufficient to study persistent compact sources near the correlation phase center, like the persistent radio counterpart to \frb. However, we also buffered the baseband EVN data to produce high-time-resolution correlations afterwards for specific times when bursts have been identified in the Arecibo single-dish data.

We used J0529+3209 as phase calibrator in all sessions ($1.1^{\circ}$ away from \frb). In the first five sessions (conducted in Feb and May) we scheduled phase referencing cycles of 8~min on the target and 1~min on the phase calibrator. Whereas this setup maximized the on-source time for burst searches, it provided less accurate astrometry due to poorer phase solutions. The pulsar B0525+21 was also observed in one of these sessions following the same strategy (phase referenced using J0521+2112), in order to perform an empirical analysis of the derived astrometry in interferometric single-burst imaging.  In the following three sessions in September, however, we conducted 5-min cycles with 3.5~min on the target and 1.5~min on the phase calibrator, improving the phase referencing, and hence providing more accurate astrometry. Two sessions failed to produce useful calibrated data on the faint target, and are not listed in Table~\ref{tab:data}. The first session (2016 Feb 1) was used to explore different calibration approaches, whereas the 2016 Sep 19 session was unusable because the largest EVN stations were unavailable and the data could not be properly calibrated without them. An extragalactic $\sim2$~mJy compact source \citep[VLA2 in][]{kon15}, was identified in the same primary beam as \frb (with coordinates $\alpha_{\rm J2000} = 5^{\rm h}31^{\rm m}53.92244^{\rm s},\ \delta_{\rm J2000} = 33\degree 10^\prime20.0739^{\prime\prime}$).  This source has been used to acquire relative astrometry of \frb during all the sessions and to provide a proper motion constraint.

The 2-s integrated data were calibrated using standard VLBI procedures within AIPS\footnote{The Astronomical Image Processing System, AIPS, is a software package produced and maintained by the National Radio Astronomy Observatory (NRAO).} and ParselTongue \citep{kettenis2006}, including {\em a priori} amplitude calibration using system temperatures and gain curves for each antenna, antenna-based delay correction and bandpass calibration. The phases were corrected by fringe-fitting the calibrators. The phase calibrator J0529+3209 was then imaged and self-calibrated using the Caltech Difmap package \citep{shepherd1994}. These corrections were interpolated and applied to \frb, which was finally imaged in Difmap.

The arrival times of the bursts were first identified using Arecibo single-dish data, and then slightly refined for application to the EVN data.  First using coherently dedispersed Arecibo auto-correlations from the EVN data, we performed a so-called gate search by creating a large number of short integrations inside a 50-ms window around the nominal Arecibo single-dish arrival times. A pulse profile was then created for each of the bursts by plotting the total power in the cross-correlations as a function of time.  We then used this pulse profile to determine the exact time window for which the correlation function was accumulated, i.e. the `gate'.  We de-dispersed and correlated the EVN data to produce visibilities for windows covering only the times of detected bursts.  We applied the previously described calibration to the single-pulse data and imaged them.  The final images were produced using a Briggs robust weighting of zero \citep{briggs1995} as it produced the most consistent results (balance between the longest baselines to Arecibo and the shorter, intra-European baselines).  Images with natural or uniform weighting did not produce satisfactory results due to the sparse $uv$-coverage.  The flux densities and positions for all datasets were measured using Difmap and CASA\footnote{The Common Astronomy Software Applications, CASA, is software produced and maintained by the NRAO.} by fitting a circular Gaussian component to the detected source in the $uv$-plane.

\section{Results} \label{sec:results}

\subsection{Burst Detections}

The EVN observations detect the compact and persistent source found by \citet{chatterjee2017} with a synthesized beam size (FWHM) of $\approx 21~\mathrm{mas} \times 2~\mathrm{mas}$ at 1.7~GHz and $\approx 4~\mathrm{mas} \times 1~\mathrm{mas}$ at 5~GHz, with position angle $\approx - 55\degree$ in both cases.

\begin{figure}[t!]
	\epsscale{1.19}
	\plotone{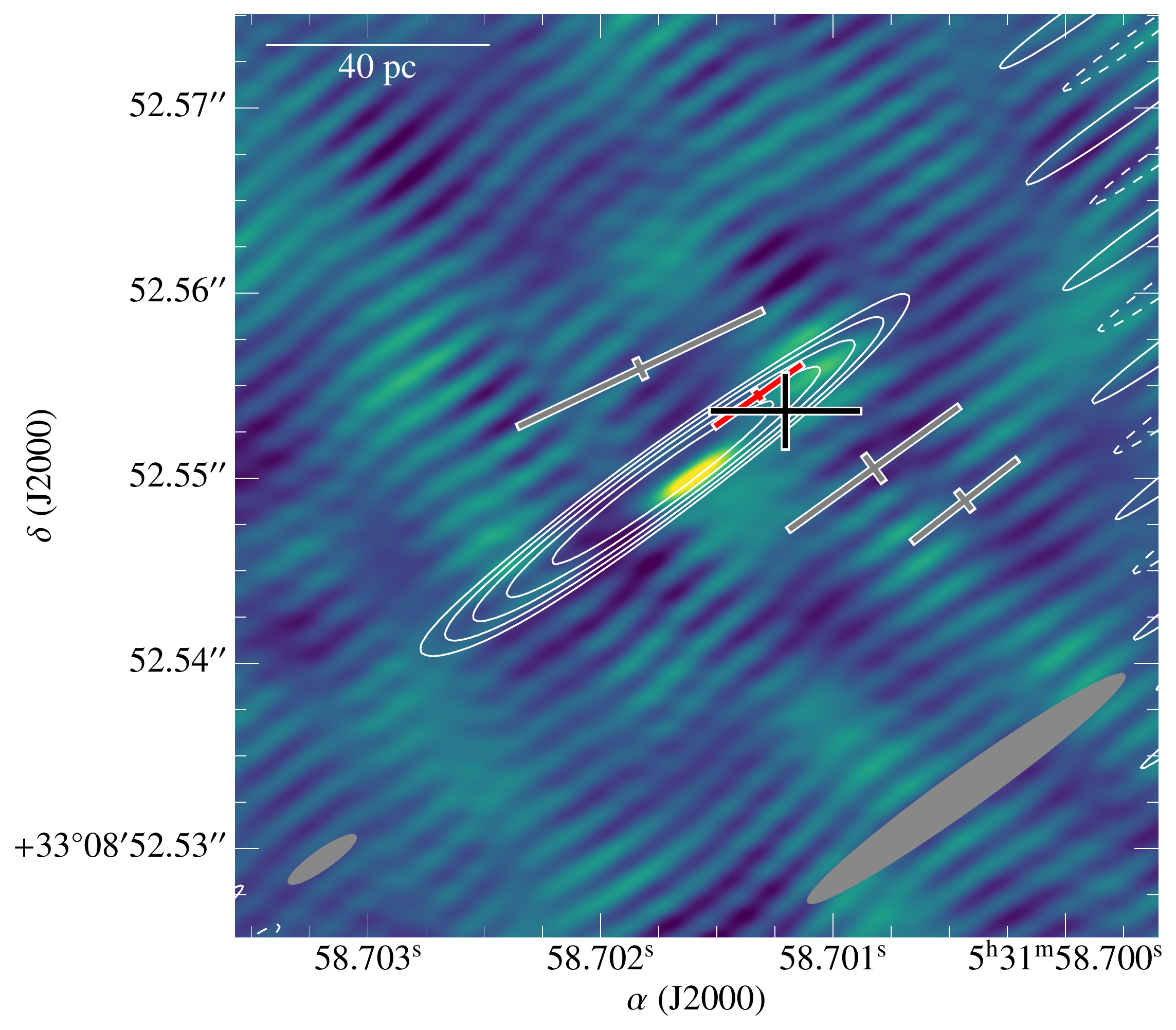}
	\caption{EVN image of the persistent source at 1.7~GHz (white contours) together with the localization of the strongest burst (red cross), the other three observed bursts (gray crosses), and the position obtained after averaging all four bursts detected on 2016 Sep 20 (black cross).  Contours start at a 2-$\sigma$ noise level of $10~\mathrm{\upmu Jy}$ and increase by factors of $2^{1/2}$.  Dashed contours represent negative levels. The color scale shows the image at 5.0~GHz from 2016 Sep 21. The synthesized beam at 5.0~GHz is represented by the gray ellipse at the bottom left of the figure and for 1.7~GHz at the bottom right.  The lengths of the crosses represent the 1-$\sigma$ uncertainty in each direction.  
    Crosses for each individual burst reflect only the statistical errors derived from their S/N and the beam size.  The size of the cross for the mean position is determined from the spread of the individual burst locations, weighted by $\xi$ (see text), and is consistent with the centroid of the persistent source to within $< 2\sigma$.
	\label{fig:burst_image}}
\end{figure}

On 2016 Sep 20, we detected four individual bursts in the Arecibo single-dish PUPPI data that overlap with EVN data acquisition (Table~\ref{tab:data}).  No bursts were detected in the Arecibo PUPPI (1.7~GHz) or Mock (5~GHz) data from other sessions in which there are simultaneous EVN observations that can be used for imaging the bursts.  We formed images from the calibrated visibility data for each burst, and measured their positions with respect to the persistent radio source. Figure~\ref{fig:burst_image} shows these positions together with the persistent source at 1.7 and 5.0~GHz. The nominal positions measured for the four bursts are spread $\lesssim 15~\mathrm{mas}$ around the position of the persistent source, and we discuss this scatter in \S\ref{sec:astrometry}.

\begin{figure}[ht!]
	\epsscale{1.15}
	\plotone{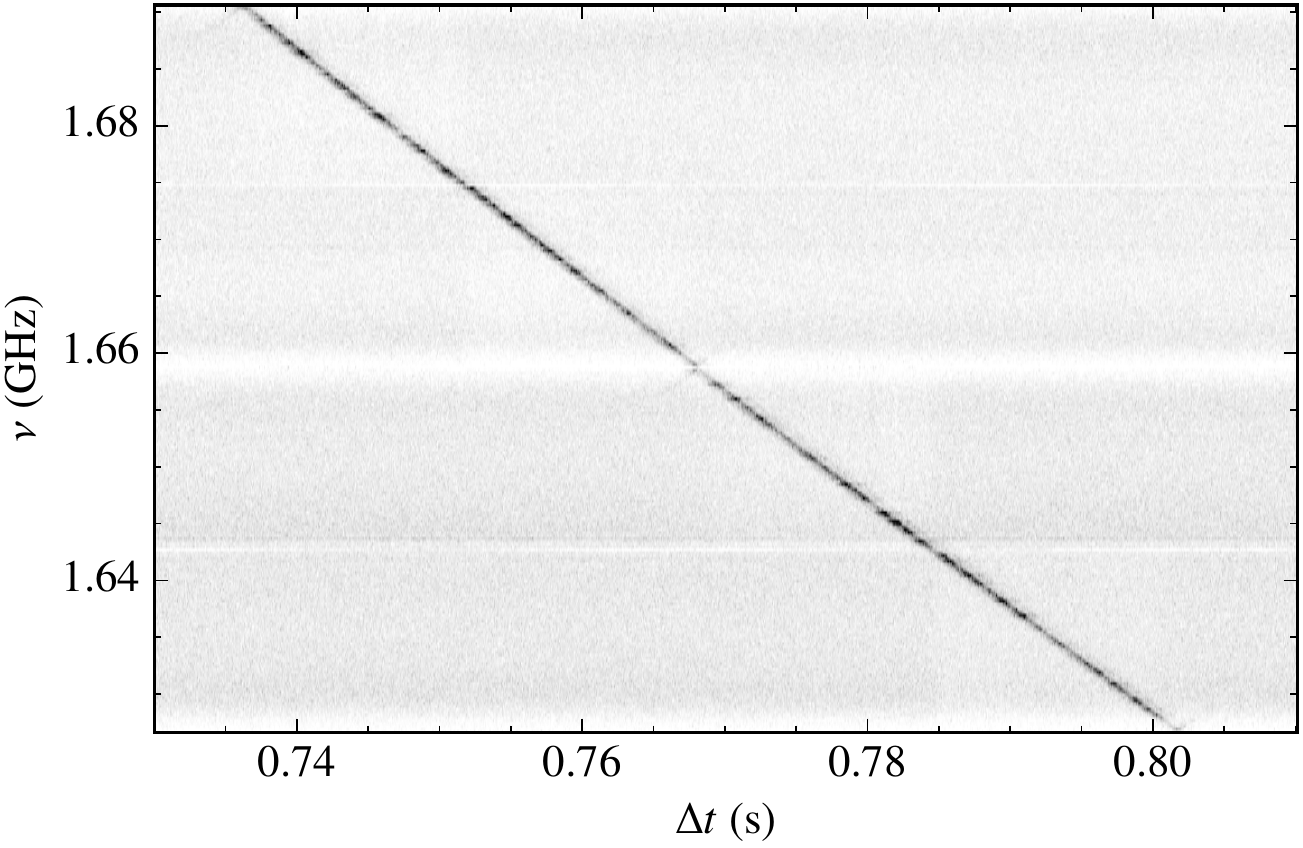}\\[+10pt]
	\plotone{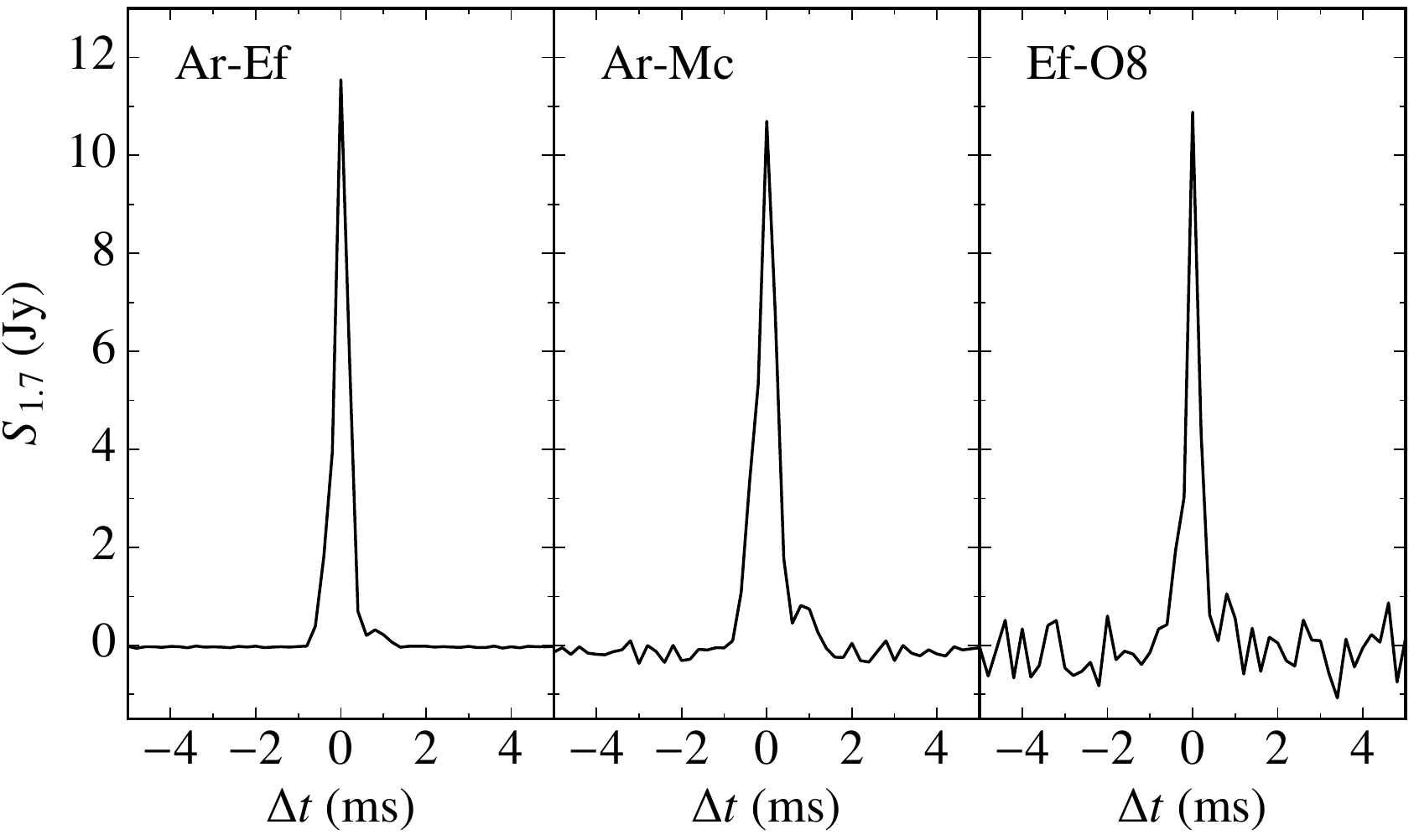}\\
    	\caption{Top: Dynamic spectrum of the strongest burst detected on 2016 Sep 20 (Burst \#2 in Table~\ref{tab:data}) from Arecibo autocorrelations, showing the dispersive sweep across the observing band.  Bottom: Coherently dedispersed and band-integrated profiles of the same burst as observed in the cross-correlations for Arecibo-Effelsberg (Ar-Ef), Arecibo-Medicina (Ar-Mc), and Effelsberg-Onsala (Ef-O8) after only applying {\it a priori} calibration. The measured peak brightnesses are 11.9, 10.7, and 10.9~$\mathrm{Jy}$, respectively, where the error is typically $10-20$\% for {\it a priori} calibration.  The rms on each baseline is 12, 80, and 300~$\mathrm{mJy}$, respectively.
	\label{fig:pulse_det1}}
\end{figure}

\begin{figure}[t!]
	\epsscale{1.15}
	\plotone{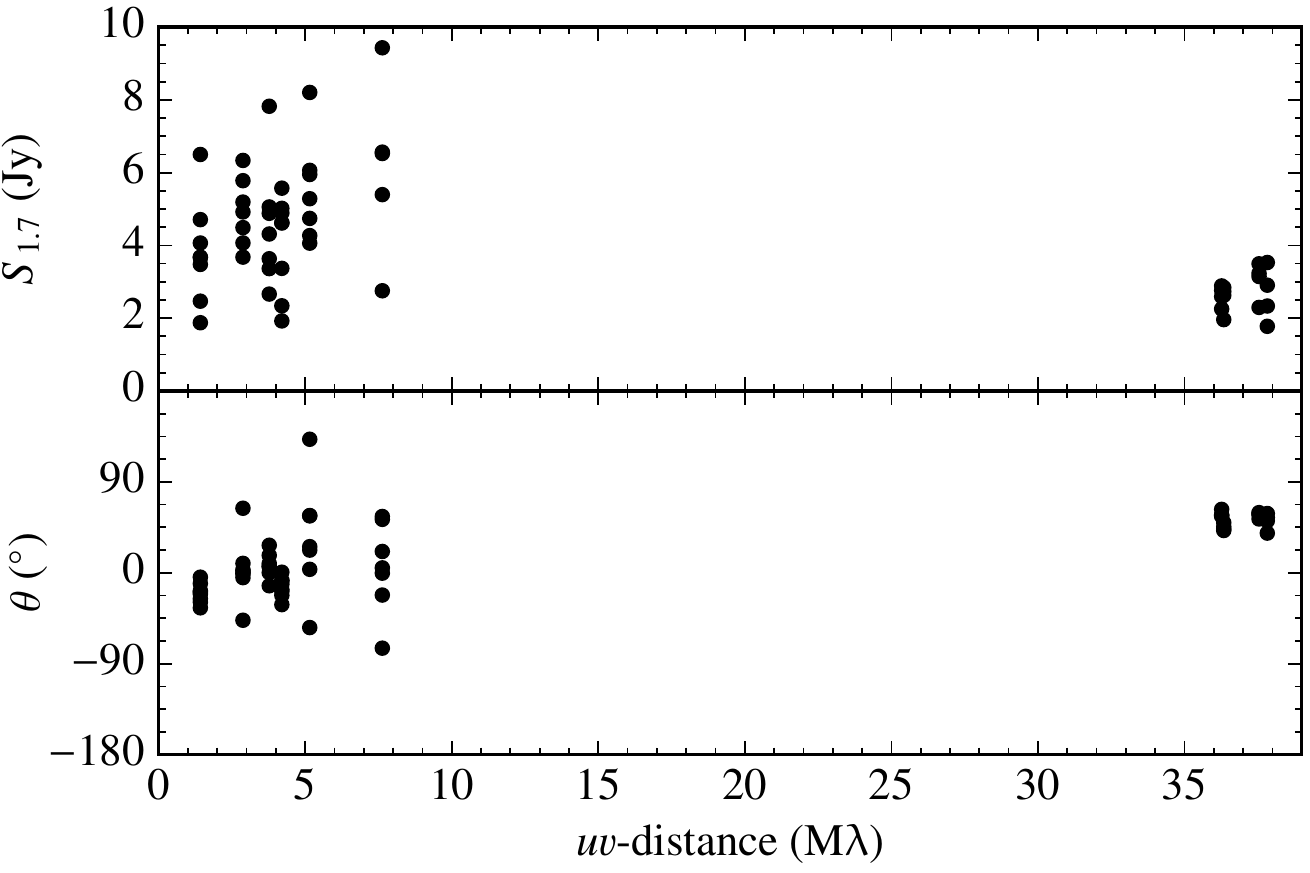}\\
	\epsscale{1.18}
	\plotone{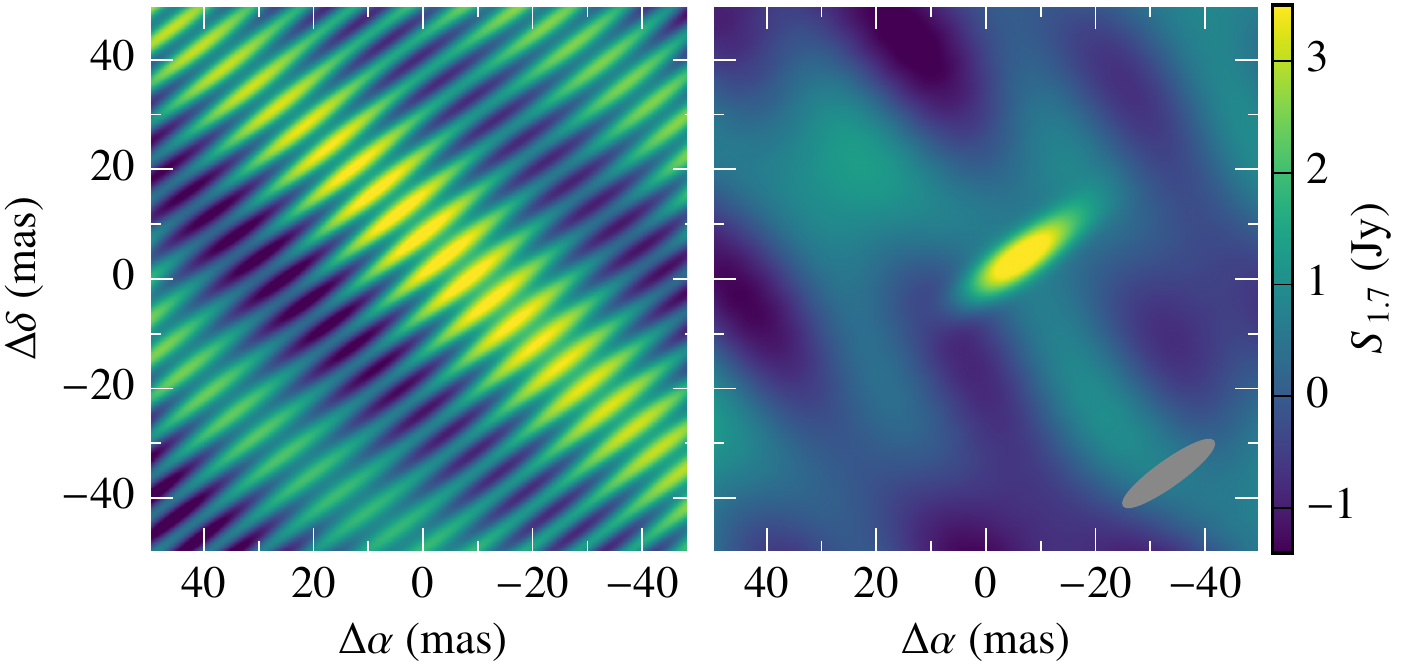}
    	\caption{Top: Amplitudes and phases of the obtained visibilities for the strongest burst observed on 2016 Sep 20 (Burst \#2 in Table~\ref{tab:data}) as a function of the $uv$-distance.
	Bottom: Dirty (left) and cleaned (right) image for the same burst. The cleaned image has been obtained by fitting the $uv$-data with a circular Gaussian component. The synthesized beam is shown by the gray ellipse at the bottom right of the figure. The coordinates are relative to the position of the persistent source obtained in the same epoch.
	\label{fig:pulse_det2}}
\end{figure}

Figures~\ref{fig:pulse_det1} and \ref{fig:pulse_det2} show data corresponding to the strongest burst (Burst~\#2) -- in the time-domain and in the image plane, respectively.  We have characterized the bursts using the detection statistic $\xi = F/\sqrt{w}$ \citep[fluence divided by the square-root of the burst width; e.g.,][]{cm03,ttw+13}.  When matched filtering is done to detect a pulse (as we have done, starting with the single-dish PUPPI data), then the S/N of the detection statistic, i.e. the output of the correlation, is proportional to $\xi$.  Localization of the source in an image (whether in the image or in the {\it uv} domain) will tend to have the same scaling if the {\it uv} data are calculated with a tight gate (time window) around the pulse so that it also scales as $w$.  Using only fluence as a detection statistic is not appropriate because a high-fluence, very wide burst can still be buried in the noise, whereas a narrower burst with equivalent fluence is more easily discriminated from noise.  Burst~\#2 was roughly an order-of-magnitude brighter than the other 3 bursts, and shows a detection statistic $\xi$ that is also a factor of $> 6$ higher than the other bursts.  This brightest burst is separated by only $\sim 7~\mathrm{mas}$ from the centroid of the persistent source at the same epoch and is positionally consistent at the $\sim 2$-$\sigma$ level.  We thus find no convincing evidence that there is a significant offset between the source of the bursts and the persistent source.  Since Burst~\#2's detection statistic, $\xi$, is significantly larger than for any of the other three bursts, its apparent position is least affected by noise in the image plane, as we explain in the following section, \S\ref{sec:astrometry}.  As such, in principle it provides the most accurate position of all 4 detected bursts, and the strongest constraint on the maximum offset between bursts and the compact, persistent radio source.

\subsection{Astrometric Accuracy}
\label{sec:astrometry}

The astrometric accuracy of full-track (horizon-to-horizon observations) EVN phase-referencing is usually limited by systematic errors due to the poorly modeled troposphere, ionosphere and other factors. These errors are less than a milliarcsecond in ideal cases \citep{pradel2006}, but in practice they can be a few milliarcseconds. Given the short duration of the bursts (a few milliseconds), our interferometric EVN data only contain a limited number of visibilities for each burst, which results in a limited $uv$-coverage and thus very strong, nearly equal-power sidelobes in the image plane (see Figure~\ref{fig:pulse_det2}, bottom panel).  In this case we are no longer limited only by the low-level systematics described above. The errors in the visibilities, either systematic or due to thermal noise, may lead to large and non-Gaussian uncertainties in the position, especially for low S/N, because the response function has many sidelobes. It is not straightforward to derive the astrometric errors for data with just a few-milliseconds integration.  Therefore, we conducted the following procedure to verify the validity of the observed positions and to estimate the errors.

First, we independently estimated the approximate position of the strongest burst by fringe-fitting the burst data and using only the residual delays (delay mapping; \citealt[in prep.]{huang2017}). With this method we have obtained an approximate position of $\alpha_{\rm J2000} = 5^{\rm h}31^{\rm m}58.698^{\rm s} (_{-0.006}^{+0.004}),\  \delta_{\rm J2000} = 33^{\circ}8^{\prime}52.586^{\prime\prime} (_{-0.044}^{+0.040})$, where the quoted errors are at the 3-$\sigma$ level. This method provides additional confidence that the image-plane detection of the bursts is genuine, since the positions obtained with the two methods are consistent at the 3-$\sigma$ level.

\begin{figure}[ht!]
	\epsscale{1.15}
	\plotone{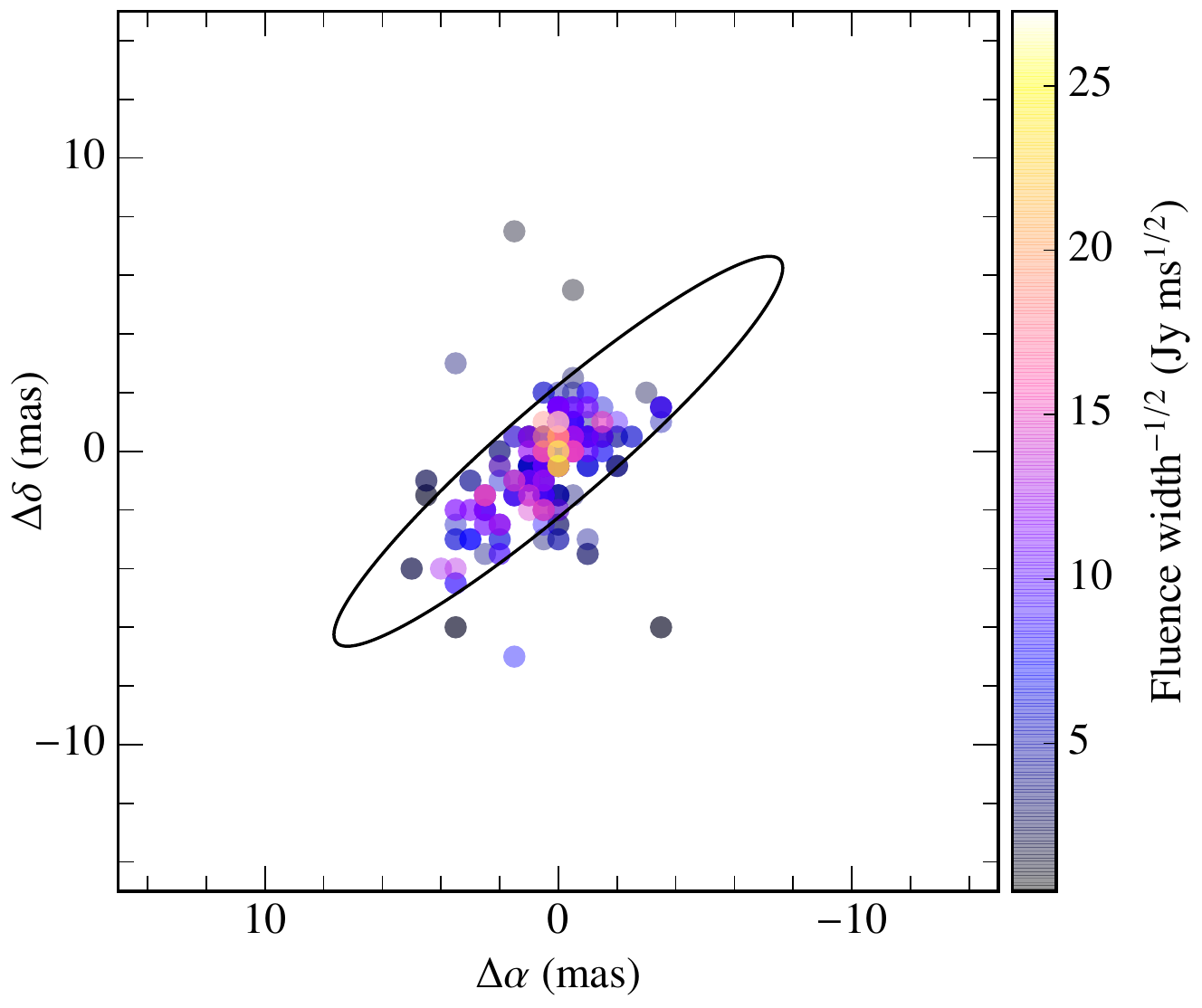}
	\caption{Pulse localizations from the pulsar B0525+21 observed on 2016 Feb 11 at 1.7~GHz. A total of 406 pulses were imaged.  Systematic uncertainties inversely proportional to the detection statistic $\xi$ are observed. We note that only pulses with $\xi \gtrsim 5$~Jy~ms$^{1/2}$ are robustly localized to within the FWHM, and for pulses with $\xi \sim 0.5$--$1$~Jy~ms$^{1/2}$ the scatter can be closer to 10~mas.
	\label{fig:test_source}}
\end{figure}

Next, we carried out an empirical analysis of single-burst EVN astrometry by imaging 406 pulses recorded from the pulsar B0525+21, which was used as a test source in the 2016 Feb 11 session.  PSR~B0525+21 has typical pulse widths of roughly 200~ms and peak flux densities of $\sim 70$--$900~\mathrm{mJy}$.  This corresponds to a range of measured detection statistics $\xi \sim 0.5$--$27$~Jy~ms$^{1/2}$, compared to the range $\xi \sim 0.2$--$5$~Jy~ms$^{1/2}$ measured for the 4 detected \frb bursts.  Figure~\ref{fig:test_source} shows the obtained positions for the different PSR~B0525+21 pulses along with the synthesized beam FWHM for comparison.  This demonstrates that the positional accuracy of the bursts increases for larger $\xi$.  It shows that pulses with $\xi \gtrsim 5$~Jy~ms$^{1/2}$ are typically offset by less than the beam FWHM, whereas for $\xi \sim 0.5$--$1$~Jy~ms$^{1/2}$ the scatter can be closer to 10~mas.  This matches well with what we have observed in the four detected \frb bursts (Table~\ref{tab:data}, Figure~\ref{fig:burst_image}).

While Burst~\#2 is thus expected to provide the most accurate position, a more conservative way to estimate the positional error on the burst source is to consider the scatter in all four detections, which is $\sim 10$~mas around the average position.  In Figure~\ref{fig:burst_image} we show the average position from the four observed bursts, weighted by their detection statistic $\xi$ (Table~\ref{tab:data}). This average position (separated $\sim 8~\mathrm{mas}$ with respect to the persistent source) shows that the average burst position and the persistent source position are coincident within $2\sigma$.  We therefore claim no significant positional offset between the persistent radio source and the source of the \frb bursts.

Finally, we place limits on the angular separation between the source of the bursts and the persistent radio source by sampling from Gaussian distributions with centers and widths given by the source positions and uncertainties listed in Table~\ref{tab:data} and deriving a numerical distribution of offsets.  Using the average burst position compared to that of the persistent source, this results in a separation of $\lesssim 12$~mas ($\lesssim 40$~pc) at the 95\% confidence level (or $\lesssim 50$~pc at 99.5\% confidence level).  Although the positional uncertainties on individual bursts are likely underestimated and non-Gaussian, as discussed previously, the effect of this should be mitigated somewhat by using the average burst position, which includes an uncertainty determined by the scatter in the separate burst detections, as also seen in Figure~\ref{fig:test_source} for B0525+21. We note that nearly identical separation limits are obtained if we consider instead the position of only the strongest burst, Burst~\#2.

\subsection{Measured Properties}

Fitting the $uv$-plane data with a circular Gaussian component shows that both the bursts and persistent radio source appear to be slightly extended. We measure a source size of $\sim 2 \pm 1~\mathrm{mas}$ at 1.7~GHz in the detected bursts in the $uv$-plane. In the persistent source we measure a similar value of $2$--$4~\mathrm{mas}$ at 1.7~GHz in all sessions, whereas at 5.0~GHz we measure an angular size of $\sim 0.2$--$0.4~\mathrm{mas}$. Measurements in the image plane (after deconvolving the synthesized beam) result in similar values. The measured source sizes for the persistent source are consistent with the ones obtained in \citet{chatterjee2017}.

\begin{figure}[t!]
	\epsscale{1.15}
	\plotone{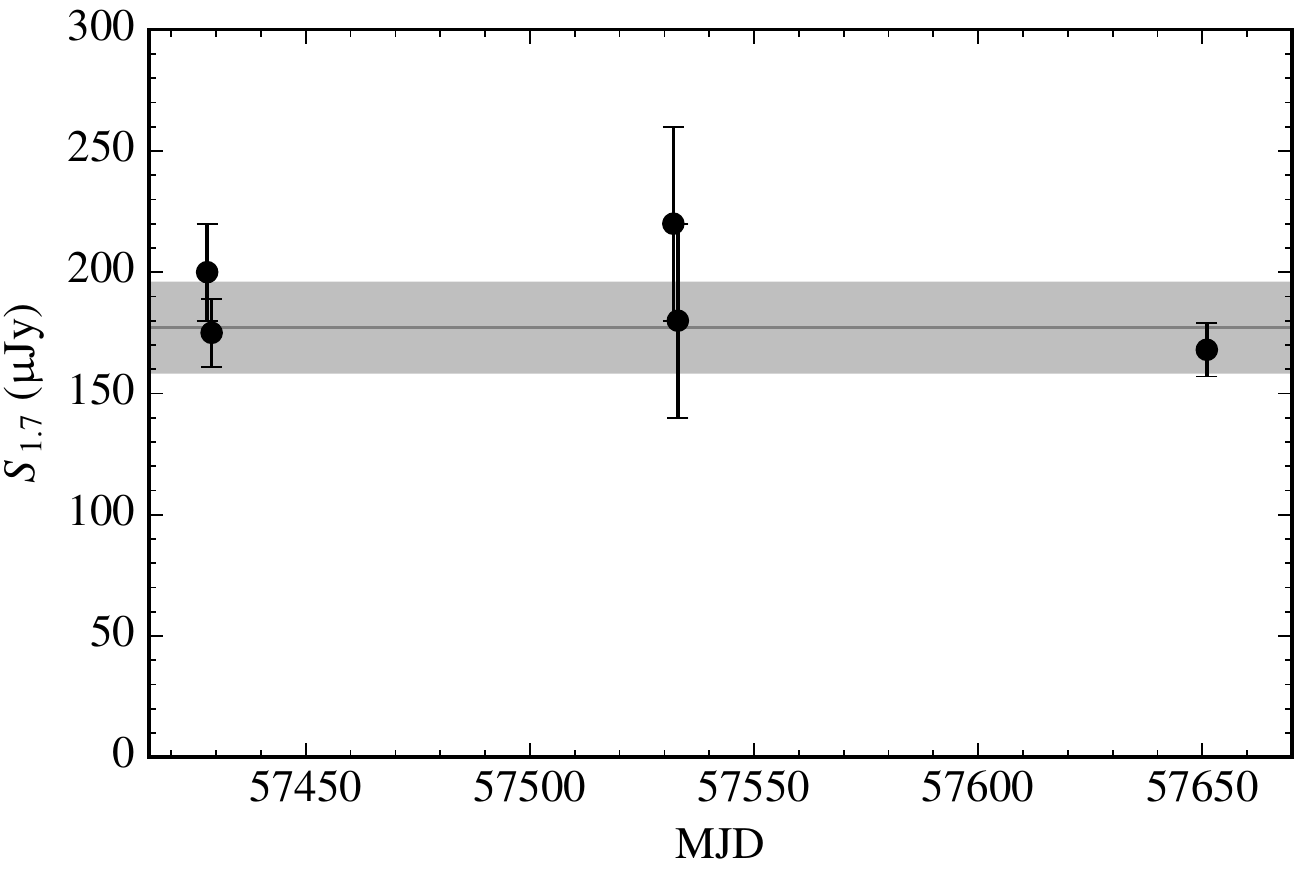}\\
	\plotone{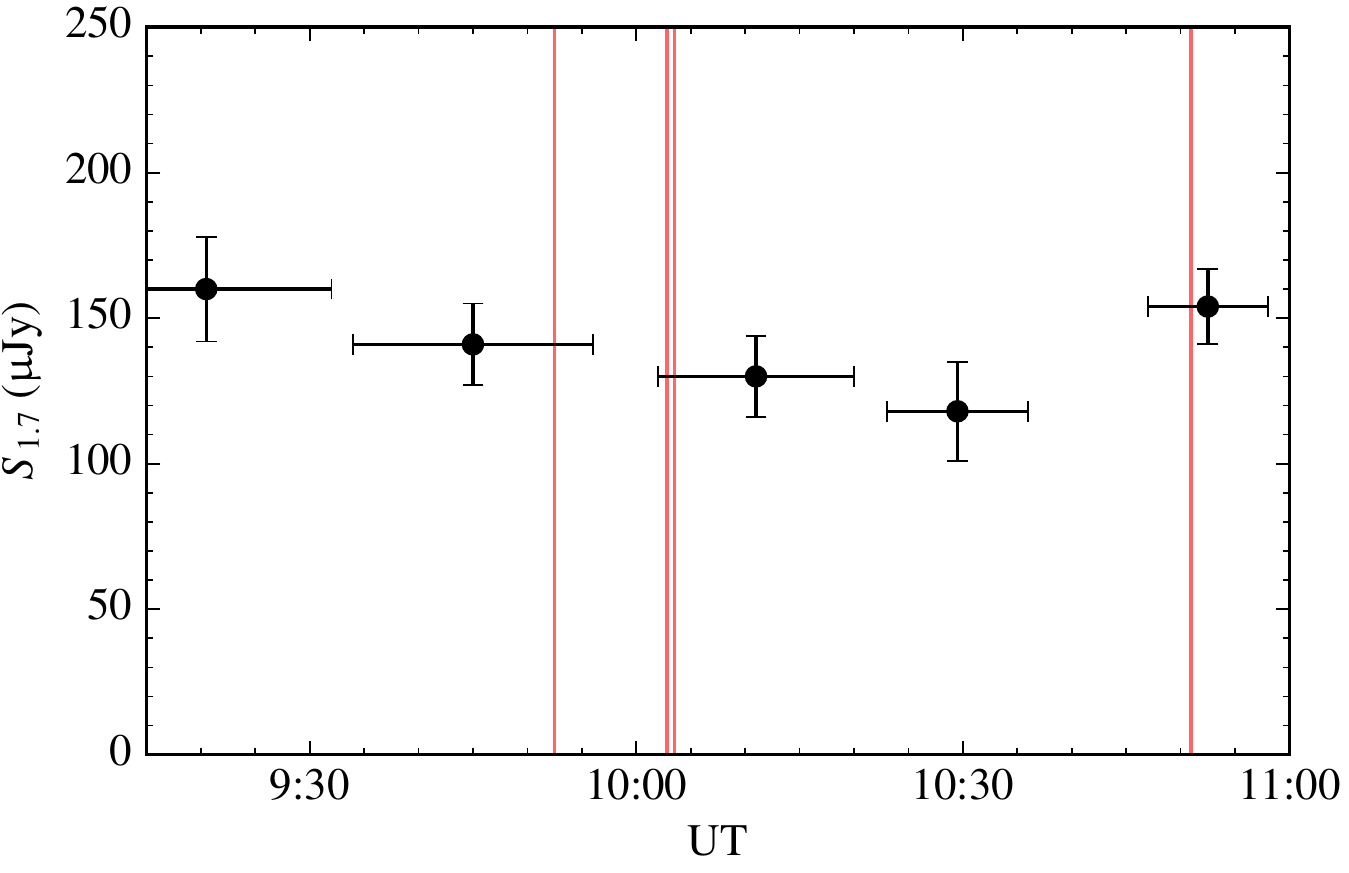}
	\caption{Top: Light curve of the persistent source at 1.7~GHz during all the EVN epochs. The horizontal line represents the average flux density value and its 1-$\sigma$ standard deviation. Bottom: Light curve of the source within the 2016 Sep 20 epoch (last epoch in the top figure). The vertical red lines represent the arrival times of the four detected bursts. We do not detect brightening of the persistent source on these timescales after the bursts.
	\label{fig:lightcurve}}
\end{figure}

Notably, Burst~\#2 shows two pulse components, separated by approximately 1~ms (Figure~\ref{fig:pulse_det1}).  Complex profile structure is commonly seen in the brightest \frb\ bursts observed to date, many of which show two or more pulse components \citep[][Hessels et al., in prep.]{ssh+16a}.  Furthermore, as can be seen in the dynamic spectrum of the burst, there is fine-scale frequency structure ($\sim$~MHz) in the intensity, which in principle could be due to scintillation or self noise.  This will be investigated in more detail in a forthcoming paper.

The different epochs at which the persistent radio source was observed allow us to obtain the light-curve of the compact source. Figure~\ref{fig:lightcurve} shows the flux densities measured for the five sessions at 1.7~GHz, which are compatible with an average flux density of $S_{1.7} = 177 \pm 18\ \mathrm{\upmu Jy}$. The only session at 5.0~GHz shows a compact source with a flux density of $S_{5} = 123 \pm 14\ \mathrm{\upmu Jy}$. Assuming that the source exhibited a similar flux density at 1.7~GHz compared with the day before, we infer a two-point spectral index $\alpha = -0.27 \pm 0.24$, where $S_{\nu} \propto \nu^{\alpha}$, for the source. Table~\ref{tab:data} summarizes the obtained results.

\section{Discussion} \label{sec:discussion}

\citet{chatterjee2017} have shown that the persistent radio source is associated with an optical counterpart, which \citet{tendulkar2017} show is a low-metallicity, star forming dwarf galaxy at a redshift of $z = 0.19273 \pm 0.00008$.
In the following, we use the luminosity and angular diameter distances of $D_{\rm L} \approx 972\ \mathrm{Mpc}$ and $D_{\rm A} \approx 683\ \mathrm{Mpc}$, respectively, determined by \citet{tendulkar2017}. We show that the VLBI data alone provide further support to the extragalactic origin of both radio sources. Furthermore, we argue that the bursts and the persistent radio source must be physically related because of their close proximity to each other.  We assume such a direct physical link in the following discussion.

\subsection{Persistent Source and Burst Properties} \label{subsec:persistent}

The results from all the EVN observations conducted at 1.7~GHz show a compact source with a persistent emission of $\sim 180\ \mathrm{\upmu Jy}$, which is consistent with the flux densities inferred at $\sim 100\times$ larger angular scales with the VLA \citep{chatterjee2017}. No significant, short-term changes in the flux density are observed after the arrival of the bursts or otherwise (Figure~\ref{fig:lightcurve}). The average flux density of the persistent source implies a radio luminosity of $L_{1.7} \approx 3 \times 10^{38}\ \mathrm{erg\ s^{-1}}$. The single measured flux density at 5.0~GHz corresponds to a similar luminosity of $L_{5.0} \approx 7 \times 10^{38}\ \mathrm{erg\ s^{-1}}$ ($\nu L_{\nu}$, with a bandwidth of 128~MHz at both frequencies). Additionally, the 5.0-GHz data allow us to set a constraint on the brightness temperature of the persistent source of $T_{\rm b} \gtrsim 5 \times 10^7\ \mathrm{K}$.  Considering the measured radio luminosities and the current 5-$\sigma$ X-ray upper limit in the $0.5$--$10$~keV band of $5 \times 10^{-15}\ \mathrm{erg\ s^{-1}\ cm^{-2}}$ \citep[which implies $L_X < 5 \times 10^{41}\ \mathrm{erg\ s^{-1}}$]{chatterjee2017} we infer a ratio between the 5.0-GHz radio and X-ray luminosities of $\log R_{\rm X} > -2.4$, where $R_{\rm X} = \nu L_{\nu} (\mathrm{5~GHz}) / L_X (\mathrm{2{\rm -}10~keV})$ as defined by \citet{terashima2003}.  The strongest observed burst exhibits a luminosity of $\sim 6 \times 10^{42}\ \mathrm{erg\ s^{-1}}$ at 1.7~GHz in the 2-ms integrated data. These values imply an energy of $\sim 10^{40}(\Delta\Omega/ 4\pi)~\mathrm{erg}$, where $\Delta\Omega$ is the emission solid angle.

With the EVN sessions spanning a period of approximately 7 months, we derived a constraint on the proper motion of the persistent source of $-6.4 < \mu_{\alpha} < 1.4\ \mathrm{mas\ yr^{-1}}$, and $-2.8 < \mu_{\delta} < 6.2\ \mathrm{mas\ yr^{-1}}$ at a 3-$\sigma$ confidence level.  These values have been obtained after removing the offsets measured in the in-beam calibrator source (VLA2, see \S~\ref{sec:obs}).  Since most of our short observations were not ideal for astrometry, this is a preliminary result, to be further improved on by follow-up observations, which will also have the advantage of spanning a longer period of time.  Nonetheless, these results already rule out the presence of parallax $\gtrsim 3~\mathrm{mas}$ at a 3-$\sigma$ confidence level, setting a distance for the persistent source $\gtrsim 0.3~\mathrm{kpc}$.

The compactness of the source at 5.0~GHz allows us to set an upper limit on its projected physical size of $\lesssim 0.7~\mathrm{pc}$ ($1.4 \times 10^5~\mathrm{AU}$, given the distance of the source).  The angular size measured for the source at 1.7 and 5.0~GHz ($\sim 2$ and $\sim 0.2$~mas, respectively) is consistent with the angular broadening expected in the direction to \frb for extragalactic sources due to local scattering (multi-path propagation) of the signal by the intervening Galactic material between the source and the observer (predicted by the NE2001 model to be $\sim 2\,\mathrm{mas}$ at 1.7~GHz; \citealt{cl02}).  Angular broadening scales as $\propto \nu^{-2}$ and thus we would expect a size of $\sim 0.4~\mathrm{mas}$ at 5.0~GHz, also roughly consistent with the measured value at that frequency.  

The obtained angular sizes are thus likely to be produced by angular broadening and not by the fact that we are resolving the source.  The angular broadening measured in the bursts ($\sim 2~\mathrm{mas}$) supports this statement as this broadening must be produced extrinsically to the bursts, given that their millisecond duration implies that the emitting region must be smaller than the light-crossing time, i.e. $\lesssim 1\,000\ \mathrm{km}$, and thus must appear to be point-like.  A caveat here is that the measured source size depends strongly on the gain of the telescope providing the longest baselines \citep{natarajan2016}, in our case Arecibo at 1.7~GHz. We exclude the possibility that the Arecibo baselines have a lower amplitude (due to a large gain error), because the measured persistent source size agrees with that obtained with the VLBA independently \citep{chatterjee2017}. At 5.0~GHz the presence of three telescopes with similar baseline lengths (Arecibo, Hartebeesthoek and Tianma) also assures the consistency of the measured source size.

The intrinsic source size of the persistent source could be as small as $7\,\mathrm{\upmu as}\ \nu_1^{-5/4}$, the limit implied by synchrotron self-absorption for a frequency of optical depth unity $\nu_1 \sim 1~\mathrm{GHz}$ \citep{harris1970}. Or if $T_{\rm b} \lesssim 10^{12}$~K by synchrotron self-Compton radiation, the size is $\gtrsim 20~\mathrm{\upmu as}\ \nu^{-1}$. These angles are too small to resolve with VLBI but could be probed with interstellar scintillations.

We have constrained the projected separation between the source of the bursts and the persistent radio source to be $\lesssim 40~\mathrm{pc}$.  Such a close proximity strongly suggests that there is a direct physical link between the bursts and the persistent source, as we now discuss in more detail.

\subsection{Possible Origins of \frb}

The data presented here, in addition to the results presented by \citet{chatterjee2017} and \citet{tendulkar2017}, allow us to constrain the possible physical scenarios for the origin of \frb.  While the fact that the bursts are located within $\lesssim 40~\mathrm{pc}$ of the persistent radio source strongly suggests a direct physical link, the persistent radio source and the source of the \frb bursts don't necessarily have to be the same object.  We primarily consider two classes of models that could explain \frb and its multiwavelength counterparts: a highly energetic, extragalactic neutron star in a young supernova remnant (SNR) or an active galactic nucleus (AGN; or analogously a black hole related system with a jet).

\subsubsection{Young neutron star and nebula}

As previously shown by \citet{ssh+16a}, the repeatability of \frb rules out an origin in a cataclysmic event that destroyed the progenitor source, e.g. the collapse of a supramassive neutron star \citep{fr14}.  The repetition and energetics of the bursts from FRB 121102 have been used to argue that it comes from a young neutron star or magnetar \citep{cw16,lbp16,pp16}. At birth, the rapid spin of such (potentially highly magnetized) objects can power a luminous nebula from the region evacuated by its SNR.

The measured luminosity for the persistent radio source cannot be explained by a single SNR or a pulsar wind nebula similar to those discovered thus far in our Galaxy. A direct comparison with one of the brightest SNRs known, Cas~A \citep[300~yr old;][]{baars1977,reed1995}, shows that we would expect an emission which is $\sim 4$ orders of magnitude fainter at the given distance of $D_{\rm L} \approx 972\ \mathrm{Mpc}$. In the case of the Crab Nebula, the expected flux density would be even fainter ($\sim 0.5~\mathrm{nJy}$) if placed in the host galaxy of \frb.  Compact star forming regions, such as seen in Arp~220, have collections of SNRs that have a luminosity and $T_{\rm b}$ consistent with the persistent radio source. However, neither the SFR of $\sim 240$--$1\,000~\msun\ \mathrm{yr^{-1}}$ nor the size of the region of $250$--$360~\mathrm{pc}$ of Arp~220 \citep{anantharamaiah2000} agrees with the properties of the persistent source associated with \frb ($0.4~\msun\ \mathrm{yr^{-1}}$ and $\lesssim 0.7~\mathrm{pc}$). 

\citet{mkm16} and \citet{pir16} discuss the properties of a young ($< 1000~\mathrm{yr}$) SNR that is powered by the spin-down power of a neutron star or white dwarf. The SNR expands into the surrounding medium and evacuates an ionized region that can be seen as a luminous synchrotron nebula. This model is also constrained by the observation that radio bursts of \frb are not absorbed by the nebula and that its DM has not evolved significantly over the last few years.
Considering all these effects, in this scenario \frb is likely to be between 100 and 1000~yr old and the persistent radio source is powered by the spin down of a rapidly rotating pulsar or magnetar.  In this case, the previously shown persistent radio source variability \citep[][where the higher cadence of observations allowed variability to be studied in more detail]{chatterjee2017} could be induced by scintillation, which is consistent with the compact (sub-milliarcsecond) SNR size at this age.  \citet{lunnan2014} and \citet{perley2016} show that superluminous supernovae (SLSNe) are typically hosted by low-metallicity, low-mass galaxies, and are possibly powered by millisecond magnetars. Additionally, it is shown that SLSNe and long gamma-ray bursts (LGRBs) could share similar environments. We note that these conditions are in agreement with the optical galaxy associated with \frb \citep{tendulkar2017}.

\subsubsection{Active galactic nucleus / accreting black hole}

Models have been proposed in which the bursts are due to strong plasma turbulence excited by the relativistic jet of an AGN \citep{romero2016} or due to synchrotron maser activity from an AGN \citep{ghi16}.  It is also conceivable to have an extremely young and energetic pulsar and/or magnetar near to an AGN \citep{pc15,cw16} -- either interacting or not. 

The persistent radio source is offset by $\sim 0.2$~arcsec (0.7~kpc) from the apparent center of the optical emission of the dwarf galaxy \citep{tendulkar2017}. Therefore, it is not completely clear whether the radio source can be associated with the galactic nucleus or not, but an offset AGN is plausible as reported in other galaxies \citep{barth2008}.
If the persistent source is indeed an AGN in the right accretion state, we can infer the mass of the black hole assuming the Fundamental Plane of black hole activity \citep{merloni2003,falcke2004,kording2006,plotkin2012,miller-jones2012}. Given the measured radio luminosity and the upper limit on the X-ray value, we estimate a lower limit on the mass of the putative black hole of $\gtrsim 2 \times 10^9\ \msun$. This value would be hard to reconcile with the fact that the stellar mass of the host galaxy is likely at least an order of magnitude less than that, and its optical spectrum shows no signatures of AGN activity \citep{tendulkar2017}.

Alternatively, we could be witnessing a radio-loud, but otherwise low-luminosity AGN powered by a much less massive black hole that accretes at a very low rate. This population is poorly known, but EVN observations of the brightest low-luminosity AGNs (LLAGNs) in a sample of Fundamental Plane outliers (i.e. radio-loud, with $R_{\rm X} \sim -2$) show that some of these have extended jets/lobes and the radio excess may come from strong interaction with the surrounding gas in the galaxy; others appear very compact like our persistent radio source, and the reason for their high $R_{\rm X}$ remains a mystery \citep{paragi2012}.  We note that there are other recent examples of LLAGNs identified based on their VLBI properties coupled with low-levels of X-ray emission and no signs of nuclear activity from the optical emission lines \citep{park2016}.  

Other possible associations, like a single X-ray binary (such as Cyg~X-3; \citealt{merloni2003,reines2011}) or an ultraluminous X-ray nebula (such as S~26 and/or IC~342~X-1; \citealt{soria2010,cseh2012}), do not fit to the measured flux density of the persistent radio emission and/or the observed size by several orders of magnitude.

\section{Conclusions} \label{sec:conclusion}

The bursts of \frb have recently been associated with a persistent and compact radio source \citep{chatterjee2017} and a low-metallicity star forming dwarf galaxy at a redshift of $z = 0.19273 \pm 0.00008$ \citep{tendulkar2017}. The EVN data presented in this work show for the first time that the bursts and the persistent source are co-located with an angular separation $\lesssim 12~\mathrm{mas}$ ($\lesssim 40~\mathrm{pc}$ given the distance to the host galaxy). This tight constraint -- roughly an order of magnitude more precise localization compared to that achieved with the VLA in \citet{chatterjee2017} -- strongly suggests a direct physical link, though the persistent radio source and the source of the \frb bursts don't necessarily have to be the same object.  Although the origin of FRBs remains unknown, the data presented here are consistent in many respects with either an interpretation in terms of a low-luminosity AGN or a young SNR powered by a highly energetic neutron star/magnetar.

\acknowledgments

We thank the directors and staff of all the EVN telescopes for making this series of target of opportunity observations possible. We thank the entire staff of the Arecibo Observatory, and in particular A.~Venkataraman, H.~Hernandez, P.~Perillat and J.~Schmelz, for their continued support and dedication to enabling observations like those presented here.  We thank B.~Stappers and M.~Mickaliger for their support with simultaneous pulsar recording using the Lovell Telescope. We thank E.~Adams, K.~Kashiyama, N.~Maddox, and E.~Quataert for useful discussions on plausible scenarios as well as O.~Wucknitz and A.~Deller for reviewing a draft of the paper. We thank F.~Camilo for access to computing resources.  The Arecibo Observatory is operated by SRI International under a cooperative agreement with the National Science Foundation (AST-1100968), and in alliance with Ana G.~M\'{e}ndez-Universidad Metropolitana, and the Universities Space Research Association. The European VLBI Network is a joint facility of independent European, African, Asian, and North American radio astronomy institutes. Scientific results from data presented in this publication are derived from the following EVN project codes: RP024 and RP026 (PI J.~Hessels).  This research made use of Astropy, a community-developed core Python package for Astronomy \citep{astropy2013} and APLpy, an open-source plotting package for Python hosted at \url{http://aplpy.github.com}.  B.M. acknowledges support by the Spanish Ministerio de Econom\'ia y Competitividad (MINECO) under grants AYA2013-47447-C3-1-P, AYA2016-76012-C3-1-P, and MDM-2014-0369 of ICCUB (Unidad de Excelencia `Mar\'ia de Maeztu').  J.W.T.H. is an NWO Vidi Fellow. J.W.T.H. and C.G.B. gratefully acknowledge funding from the European Research Council under the European Union's Seventh Framework Programme (FP/2007-2013) / ERC Grant Agreement no. 337062 (DRAGNET).  Y.H. would like to acknowledge the support of the ASTRON/JIVE International Summer Student Programme. ASTRON is an institute of the Netherlands Organisation for Scientific Research (NWO). The Joint Institute for VLBI ERIC, is a European entity established by six countries and funded by ten agencies to support the use of the European VLBI Network.  S.C., J.M.C., P.D., M.A.M., and S.M.R. are partially supported by the NANOGrav Physics Frontiers Center (NSF award 1430284). Work at Cornell (J.M.C., S.C.) was supported by NSF grants AST-1104617 and AST-1008213.  M.A.M. is supported by NSF award \#1458952.  L.G.S. gratefully acknowledges financial support by the European Research Council for the ERC Starting Grant BEACON under contract no. 279702, and the Max Planck Society.  V.M.K. holds the Lorne Trottier Chair in Astrophysics and Cosmology and a Canadian Research Chair in Observational Astrophysics and received additional support from NSERC via a Discovery Grant and Accelerator Supplement, by FQRNT via the Centre de Recherche Astrophysique de Qu\'ebec, and by the Canadian Institute for Advanced Research.  S.B.-S. is a Jansky Fellow of the National Radio Astronomy Observatory.  Part of the research was carried out at the Jet Propulsion Laboratory, California Institute of Technology, under a contract with the National Aeronautics and Space Administration.  P.S. is a Covington Fellow at the Dominion Radio Astrophysical Observatory.  S.P.T. acknowledges support from a McGill Astrophysics postdoctoral fellowship.

\facility{EVN, Arecibo Observatory}
\software{AIPS, Difmap, ParselTongue, CASA, Astropy, APLpy, PRESTO}

\end{document}